# Dual Layer Textual Message Cryptosystem with Randomized Sequence of Symmetric Key

Chandranath Adak

*Abstract*—This paper introduces a new concept of textual message encryption and decryption through a pool of randomized symmetric key and the dual layer cryptosystem with the concept of visual cryptography and steganography. A textual message is converted into two image slides, and the images are encrypted through two different randomized sequences of symmetric key. The decryption is done in the reverse way. The encrypted images are decrypted by those two symmetric keys. The decrypted image slides are merged together and converted into textual message. Here the image sharing is done through the concept of visual cryptography and the textual message to image conversion is done through the concept of steganography.

## I. INTRODUCTION

CRYPTOGRAPHY is the science of techniques to secure the communication and protect the private information from anonymous[1], i.e. the art of transforming messages to make them secure and immune to attack[2].

A key is a set of number(s) that an algorithm (cipher) operates on to encrypt a message. The cryptography algorithms (ciphers) are divided into two groups: symmetric (secret) key ciphers and asymmetric (public) key ciphers. In symmetric key cryptography, the same key is used by both parties (sender and receiver)[2]. In the prescribed cryptosystem, the symmetric key is generated randomly, i.e. each time of message passing the key is different.

Steganography is the art and science of writing hidden messages in a different form, such that no one, apart from the sender and intended recipient, suspects the existence of the message, it is a form of security through obscurity[3]. In the prescribed cryptosystem, the textual message is sent in the form of image.

Visual cryptography is a method for protecting image-based secrets that has a computation-free decoding process. Naor & Shamir[4] demonstrated a visual secret sharing scheme, where an image was broken up into *n* shares so that only someone with all *n* shares could decrypt the image, while any *n-1* shares revealed no information about the original image. In the prescribed cryptosystem, the image is broken into two slides.

## II. STEPS OF PROPOSED METHOD

### A. Key Generation

A pool of random decimal numbers is generated through a random number generating function. These decimal numbers are converted into equivalent binary numbers. This collection of zeros and ones are stored to feed as symmetric key for encryption and decryption.

### B. Encryption Methodology

The source textual message is scanned character by character and the ASCII values of each character are collected. The ASCII values are treated as decimal numbers and converted into equivalent eight bit binary numbers. The pool of zeros and ones are collected from these binary numbers. This mixture of zeros and ones are separated on the basis of their odd/even position. The odd positioned zeros and ones make image1 and even positioned zeros and ones make image2 ( by the pixel values as zero and one ).

Two different set of symmetric keys (key1 and key2) are generated to encrypt the two images (image1 and image2 respectively).

The encryption is done through the 'bitwise xor' operation between the pixel values of each image (image1 and image2) and key values (key1 and key2) to form encrypted images (image1 ^ key1 = encryptedImage1 , image2 ^ key2 = encryptedImage2 ; ^ is bitwise xor symbol).

### C. Decryption Methodology

The decryption is done through the 'bitwise xor' operation between the pixel values of each encrypted image (encryptedImage1 and encryptedImage2) and key values (key1 and key2) to form decrypted images (encryptedImage1 ^ key1 = decryptedImage1 , encryptedImage2 ^ key2 = decryptedImage2 ; ^ is bitwise xor symbol).

Mr. Chandranath Adak is doing his M.Tech. in Computer Science and Engineering from University of Kalyani, Kalyani-741235, India.
( e-mail : adak32@gmail.com ).

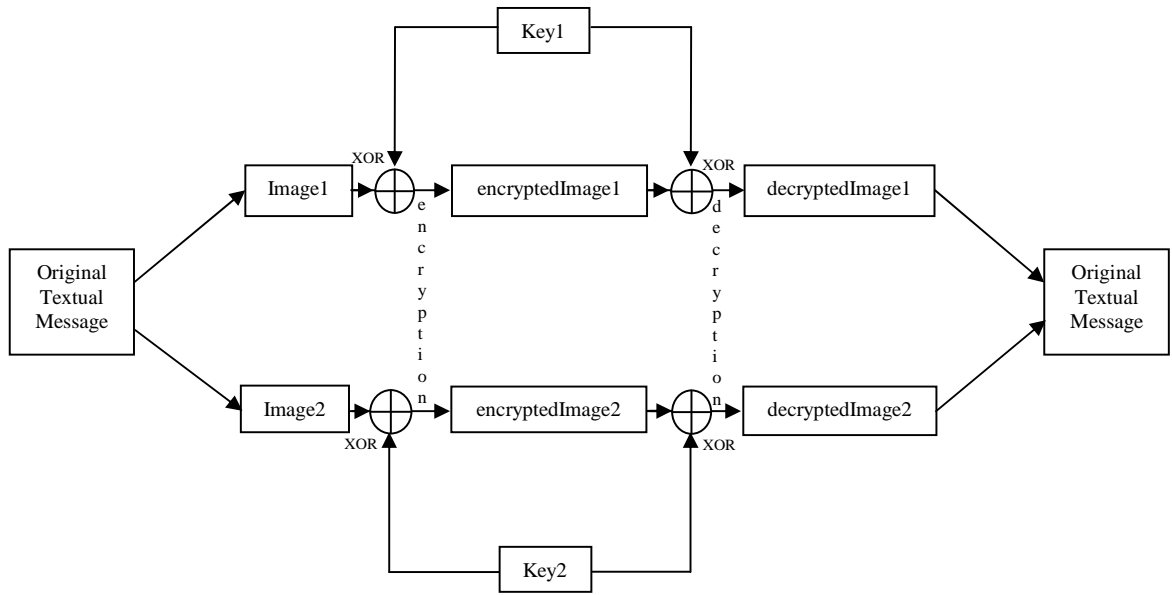

*fig. 1 :* Model of the proposed method

The decrypted images pixel values are the combination of zeros and ones. This combination of zeros-ones is collected from decryptedImage1 and decryptedImage2 pixels separately and stored in odd and even position respectively. This collection of zeros and ones are taken as stream of eight bit binary numbers and converted into equivalent decimal numbers. These decimal numbers are treated as ASCII values and each character is generated through these ASCII values. This collection of characters is the original textual message.

### III. IMPLEMENTATION

The following is the illustration of the proposed methodology with an example.

Let, the original text message contains a single letter "a". Its ASCII value is '97', which is taken as decimal number and converted into eight bit binary number. $(97)_{10} = (01100001)_2$ . This binary number is stored in the reverse way and shared in two parts with respect to odd-even position.

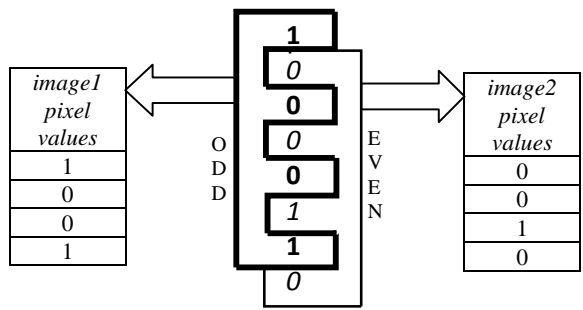

*fig. 2 :* Two slides of image formation with binary number partitioning

The odd and even positioned zeros-ones act as pixel values and form image1 and image2 respectively as fig. 2.

The symmetric keys are generated from the pool of random numbers, which is generated through some random decimal number generating function. These random decimal numbers are converted into equivalent binary numbers and stored as key values. The two symmetric keys (key1 and key2) are resized to make same-sized with the image1 and image2.

In the encryption process, the image pixel values and key values are merged with 'bitwise xor' to form the encryptedImage pixel values as table 1.

| image1 pixel values (i1) | key values (k1) | encryptedImage1 pixel values ( i1^k1) |
|---|---|---|
| 1 | 1 | 0 |
| 0 | 1 | 1 |
| 0 | 1 | 1 |
| 1 | 0 | 1 |

Table 1.(a)

| image1 pixel values (i2) | key values (k2) | encryptedImage2 pixel values ( i2^k2) |
|---|---|---|
| 0 | 0 | 0 |
| 0 | 0 | 0 |
| 1 | 0 | 1 |
| 0 | 1 | 1 |

Table 1.(b)

*Table 1.(a)* shows encryptedImage1 pixel values those are form with 'bitwise xor' operation of image1 pixel values and key1 values, similarly, *Table 1.(b)* shows encryptedImage2 pixel values.

In the decryption process, encryptedImage pixel values and key values are merged with 'bitwise xor' to form the decryptedImage pixel values as table 2.

| *encryptedImage1 pixel values (ei1)* | *key values (k1)* | *decryptedImage1 pixel values ( ei1^k1)* |
|---|---|---|
| 0 | 1 | 1 |
| 1 | 1 | 0 |
| 1 | 1 | 0 |
| 1 | 0 | 1 |

Table 2.(a)

| *encryptedImage2 pixel values (i2)* | *key values (k2)* | *decryptedImage2 pixel values ( ei2^k2)* |
|---|---|---|
| 0 | 0 | 0 |
| 0 | 0 | 0 |
| 1 | 0 | 1 |
| 1 | 1 | 0 |

Table 2.(b)

*Table 2.(a)* shows decryptedImage1 pixel values those are form with 'bitwise xor' operation of encryptedImage1 pixel values and key1 values, similarly, *Table 2.(b)* shows decryptedImage2 pixel values.

The decryptedImage1 pixel values are placed in odd position and decryptrdimage2 pixel values are placed in even position to form a sequence of zeros and ones as fig. 3. This sequence is sliced in eight bit block; each block is treated as an eight bit binary number.

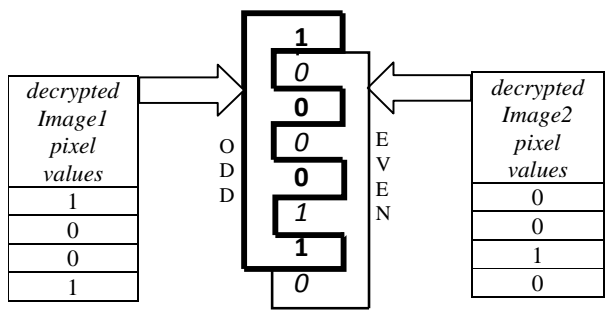

*fig. 3 :* Binary number formation with decrypted image merging

This binary number is stored in reverse way as '01100001' and converted into its equivalent decimal. $(01100001)_2 = (97)_{10}$. This decimal number is treated as ASCII value, whose equivalent character is 'a'. And original textual message was "a". So, the original textual message is recovered with the help of this prescribed method.

## IV. ADVANTAGES

The prescribed method generates random symmetric key, the key is not fixed, and it is changing in each time encryption.

The original textual message is hiding in the form of image, conception of steganography is used.

The image is shared into two parts, which is the image sharing concept of visual cryptography.

As the dual layer concept is used, the security is more. After one portion of decryption is done with the help of encryptedImage1 and key1, then the sender sends the encryptedImage2 and key2 after receiving the acknowledgement from receiver.

## V. CONCLUSION AND FUTURE WORKS

The prescribed method is the combination of steganography, visual cryptography, and a new idea about dual layer parallel cryptosystem with randomized sequence of symmetric key. This cryptosystem is tested with huge amount of data, i.e. textual messages, and result is satisfactory. Due to the use of symmetric key, there is a little bit possibility of key leakage. The next venture of this prescribed cryptosystem will be analyze the performance on the basis of parameters like computing time, execution complexity, and network overhead.

### ACKNOWLEDGMENT

I would like to heartily thank Prof. Bidyut B. Chaudhuri, Head, Computer Vision and Pattern Recognition Unit, Indian Statistical Institute, Kolkata 700108, India, for discussion various aspects of this paper.